\documentclass[apj]{emulateapj}

\usepackage{bm}

\slugcomment{Accepted by ApJ}
\shorttitle{Thermally Stable Disk}
\shortauthors{Hirose et al.}

\begin{document}

\title{Radiation-Dominated Disks Are Thermally Stable}

\author{Shigenobu Hirose}
\affil{The Earth Simulator Center, JAMSTEC, Yokohama, Kanagawa 236-0001, Japan}

\author{Julian H. Krolik}
\affil{Department of Physics and Astronomy, Johns Hopkins University, 
    Baltimore, MD 21218}

\and

\author{Omer Blaes}
\affil{Department of Physics, University of California, Santa Barbara,
   Santa Barbara CA 93106}

\begin{abstract}

When the accretion rate is more than a small fraction of Eddington, the
inner regions of accretion disks around black holes are expected to be
radiation-dominated.  However, in the $\alpha$-model, these regions are
also expected to be thermally unstable.  In this paper, we report two
3-d radiation MHD simulations of a vertically-stratified shearing box
in which the ratio of radiation to gas pressure is $\sim 10$, and yet
no thermal runaway occurs over a timespan $\simeq 40$ cooling times.
Where the time-averaged dissipation rate is greater than the critical
dissipation rate that creates hydrostatic equilibrium by diffusive
radiation flux, the time-averaged radiation flux is held to the
critical value, with the excess dissipated energy transported by
radiative advection. 
Although the stress and total pressure are well-correlated as predicted
by the $\alpha$-model, we show that stress fluctuations precede
pressure fluctuations, contrary to the usual supposition that the
pressure controls the saturation level of the magnetic energy.  This
fact explains the thermal stability.  Using a simple toy-model, we
show that independently-generated magnetic fluctuations can drive
radiation pressure fluctuations, creating a correlation between
the two while maintaining thermal stability.

\end{abstract}


\section{Introduction}

As first demonstrated by \citet{sha73}, radiation pressure must dominate gas pressure
in the innermost parts of black hole accretion disks radiating at any
non-negligible fraction of the Eddington limit.  Although the original
argument made use of the assumption that the stress and the pressure are related by
a fixed ratio $\alpha$, the critical radius within which radiation pressure exceeds
gas pressure depends so weakly on the value of $\alpha$ that this conclusion would
be virtually unaltered even if the ratio of stress to pressure varied substantially
from place to place.  In this sense, the expectation of radiation dominance in the
inner rings of brightly-radiating accretion disks rests only on the supposition
that stress and pressure are comparable, the dimensional analysis foundation
from which the $\alpha$ model began.  It is therefore crucial that we understand
the role that radiation pressure plays in the physics of accretion disks if
we are to understand luminous
black hole sources.  Yet, for the past thirty years the canonical
internal equilibrium of disks in the radiation-dominated limit has
been believed to be unstable \citep{lig74,sha76,pir78}.

In contrast to other physical regimes, the vertical structure of a stationary
disk is in some ways very tightly constrained if radiation
pressure truly dominates the hydrostatic support of the disk against the
tidal gravitational field of the hole.  Neglecting relativistic corrections
for simplicity, one finds that the outward radiation flux in a geometrically
thin disk as a function of height $z$ above the midplane is
\begin{equation}
F(z)={c\over\kappa}g(z)={c\over\kappa}\Omega^2z,
\label{eqfz}
\end{equation}
where $c$ is the speed of light, $\kappa$ is the opacity (usually dominated
by electron scattering), and the orbital angular velocity $\Omega$ determines the
vertical tidal gravitational acceleration $g$.  Because energy conservation
in a time-steady disk demands (if we neglect inner boundary condition effects)
that the flux emerging from the disk surface is $(3/8\pi)\dot M \Omega^2$, the
half-thickness $H$ of the disk depends only on the accretion rate $\dot{M}$ and
is independent of radius and the nature of the turbulent stress:
$H=3\kappa\dot{M}/(8\pi c)$.

Moreover, equation (\ref{eqfz}) combined with the condition of radiative
equilibrium immediately implies that the dissipation rate per unit volume $Q$
is constant as a function of height above the disk midplane:
$Q = dF/dz = c\Omega^2/\kappa$.  Given that $Q$
must, on average, be given by the turbulent stress
$\tau_{r\phi}$ times the rate of strain $r|d\Omega/dr|$, we also have a
tight constraint on the vertically-averaged stress:
\begin{equation}
\tau_{r\phi}={2c\Omega\over3\kappa},
\end{equation}
a conclusion first reached in slightly different form by \citet{sha76}.

On the other hand, other aspects of radiation-dominated disks are left
totally unconstrained.  Because $\kappa$ (when it is dominated by electron scattering)
is independent of density, the vertical density profile $\rho(z)$ is
completely irrelevant to both the hydrostatic and the thermal equilibrium
of the disk.  In the absence of additional information about how the dissipation
rate depends on density, it is therefore completely undetermined, and even
within Shakura-Sunyaev models
it must be specified in some {\it ad hoc} manner.  Often it is assumed that the
dissipation rate per unit mass is constant, which leads to a vertically
constant density out to the photospheres.  This element of arbitrariness
is in addition, of course, to the {\it ad hoc} ``$\alpha$-viscosity''
prescription for the stress used in such models.

The freedom to choose arbitrary values of $\alpha$ in that stress prescription
has always been one of its unfortunate aspects, but in the radiation pressure
dominated regime we also have freedom in the choice of which pressure with which
to scale the stress.  The standard \citet{sha73} assumption is that the stress
$\tau_{r\phi}$ should be proportional to the total (gas plus radiation) pressure,
and this choice leads to both thermal (e.g. \citealt{sha76}) and ``viscous''
(more properly, ``inflow"; \citealt{lig74})
instabilities where the radiation to gas pressure
ratio exceeds 3/2.  The thermal instabilities generally have faster growth rates
than the inflow modes, and are the ones of primary interest here.

The origin of the thermal instability is most simply expressed in terms
of the different dependences of the heating and cooling rates on the
midplane temperature $T$, assuming constant surface mass density $\Sigma$
\citep{pir78}.  In the limit of radial wavelength long compared to the disk
thickness, radiative diffusion implies that the cooling rate per unit area
$F^-\sim4caT^4/(3\kappa\Sigma)\propto T^4/\Sigma$, where $a$ is the radiation
density constant.  This must balance the heating rate per unit area
$F^+\sim2H\tau_{r\phi}r|d\Omega/dr|$.  Combining these two expressions with
the $\alpha$-prescription $\tau_{r\phi}=\alpha aT^4/3$ and the condition of
hydrostatic equilibrium, we find $H=2aT^4/(3\Sigma\Omega^2)$ and
$F^+\sim2\alpha a^2T^8/(3\Omega\Sigma)\propto T^8/\Sigma$.  Thermal instability
follows because positive temperature perturbations lead to greater increases in
heating than in cooling.  It is also possible to express
these relationships in terms of the total radiation energy content per
unit area $U\sim2aT^4 H$, in which case $F^-\propto U^{1/2}/\Sigma^{1/2}$ and
$F^+\propto U$; once again, instability is indicated.

However, it has never been clear whether the assumption that the
stress is proportional to the radiation pressure is correct, and the existence
of the thermal instability has therefore always been suspect.  Alternative
prescriptions that make the stress proportional to either the gas pressure
alone or some combination of the gas and radiation pressure have been
advocated over the years, and give rise to more stable
configurations (e.g. \citealt{sak81,ste84,bur85,szu90,mer02,mer03}).
Moreover, all of these prescriptions refer to vertically-integrated or
one-zone vertical structure models of the accretion disk.  The actual
vertical distribution of dissipation may also be important.  If much
of the local accretion power is actually dissipated in optically thin
regions above and below the disk, then the disk itself can become supported
by gas pressure and there would then be no thermal instability
\citep{sve94}.

It is now widely believed that the turbulence in black hole accretion disks
resembles that seen in simulations of the nonlinear development of the
magnetorotational instability (MRI, e.g. \citealt{bal98}).  Such simulations,
if they include radiation physics, could in principle help resolve these
uncertainties by calculating the spatial and temporal structure of the
turbulent dissipation within the disk, and checking to see whether a long-lived,
and therefore presumably stable, equilibrium exists in the radiation pressure
dominated regime.  It may also be possible to investigate the scaling of
average turbulent stresses with local averages of gas and/or radiation pressure
using the data from these simulations.

Well-resolved, global simulations of optically thick accretion disks that
include radiation transport have not yet been published, but it is
computationally feasible to investigate the thermal physics locally within
the disk using a stratified shearing box geometry \citep{bra95,sto96,mil00}.
The necessity of (sheared)-periodic radial boundary conditions zeroes any net
accretion rate through the box, precluding any investigation of inflow
instabilities, but thermal instabilities can still be studied.  These
same boundary conditions also preclude studying any modes with wavelengths
longer than twice the box width, except for infinite wavelength modes---the
sheared periodic radial boundary conditions place no restriction on modes
that are completely independent of radius.

The first attempt at this was made by \citet{tur04}, who performed
a radiation magnetohydrodynamic (MHD) simulation of a stratified shearing box
resulting in an average midplane radiation to gas pressure ratio of 14.
Despite this large ratio, the simulation produced no obvious
thermal instability over eight thermal times.  The robustness of this result
is somewhat suspect, however, given that the computational methods employed
precluded the photospheres (top and bottom) from being included within
the simulation domain.  Moreover, $27\%$ of the work done on the box
disappeared due to numerical losses, and half the mass was lost from the
simulation domain during a heating/expansion phase.

More recently, we \citep{hir06} developed numerical techniques that
conserve energy with high accuracy, place both photospheres within
the problem volume, and retain nearly all the initial mass on the grid.
Solving the total energy equation enables us to capture all grid scale losses
of magnetic and kinetic energy and convert them to internal energy in the gas.
Furthermore, by simultaneously solving the internal energy equation, we can
compute the instantaneous local losses of 
magnetic and kinetic energy, i.e. the energy dissipation rate.  The only
violations of energy conservation come from the small amounts of energy
(generally $\sim 0.1$\% of the total dissipated energy\footnote{0.09\% in ``1112a'' and 0.15\% in ``1126b''; see section \ref{sec:initial_condition} for the labels.})
artificially injected by the action of the density floor and similar
numerical limiters.
Using this code, we have previously studied a case in which
gas pressure dominated over radiation pressure \citep{hir06} and, in
\citet{kro07,bla07}, one in which the gas and radiation pressures were
comparable.

Both of the simulations studied in our previous papers achieved a thermal
balance between heating and cooling on time scales comparable to the
thermal time.  However, no real steady state was achieved in the simulation
with comparable gas and radiation pressure.  Instead, there were long
term ($\sim7$ thermal times) up and down variations in total energy content
by factors of three to four.  The two hottest epochs in this simulation
marginally violated the thermal instability criterion of
\citet{sha76}, but no thermal runaway occurred.

In this paper we present results of two simulations that are in the fully
radiation pressure dominated regime, with box-integrated radiation to gas
pressure ratios $\sim10$.  The two simulations shared the same parameters,
boundary conditions, and (almost) the same initial conditions, differing
only in the small noise fluctuations imposed on the initial state to seed
the MRI.  Despite being in gross violation of the
standard thermal instability criterion, there is no evidence for such
an instability in either of their time evolutions, which in both cases
extended over $\sim 40$ thermal times.

This paper is organized as follows.  In section 2 we briefly summarize
how the code works and discuss the parameters of the new simulations.
Quantitative results about the thermodynamics, internal vertical structures,
and variability of the simulations are presented in section 3.  In section 4
we discuss a simple model that might
explain the thermal stability we observe, and we summarize our
findings in section 5.


\section{Calculation}

\subsection{Basic Equations and Numerical methods}

The basic equations used in our simulations are the frequency-averaged equations
of radiation MHD in the flux-limited diffusion (FLD) approximataion:
\begin{eqnarray}
&& \frac{\partial\rho}{\partial t} + \nabla\cdot(\rho\bm{v}) = 0 \label{eq:cont}\\
&&\frac{\partial(\rho\bm{v})}{\partial t} + \nabla\cdot(\rho\bm{v}\bm{v}) =
 -\nabla(p + q) \nonumber \\
 &&\quad + \frac{1}{4\pi}(\nabla\times\bm{B})\times\bm{B} + \frac{(\bar{\kappa}_{\rm ff}^{\rm R} + \kappa_{\rm es})\rho}{c}\bm{F} + \bm{f}_{\rm SB}
\label{eq:N-S} \\
&& \frac{\partial e}{\partial t} + \nabla\cdot(e\bm{v}) = -(\nabla\cdot\bm{v})(p + q) \nonumber \\
 &&\quad - (4\pi B - cE)\bar{\kappa}_{\rm ff}^{\rm P}\rho - cE\kappa_{\rm es}\rho\frac{4k_{\rm B}(T - T_{\rm rad})}{m_{\rm e}c^{2}} \nonumber \\
 &&\quad + \tilde{Q}_{\rm diss}
\label{eq:mattenergy}\\
&& \frac{\partial E}{\partial t} + \nabla\cdot(E\bm{v}) = - \nabla\bm{v}:\mathsf{P} +  (4\pi B - cE)\bar{\kappa}_{\rm ff}^{\rm P}\rho \nonumber  \\
 &&\quad + cE\kappa_{\rm es}\rho\frac{4k_{\rm B}(T - T_{\rm rad})}{m_{\rm e}c^{2}} - \nabla\cdot\bm{F}
\label{eq:radenergy}\\
&& \frac{\partial\bm{B}}{\partial t} - \nabla\times(\bm{v}\times\bm{B}) = 0
\label{eq:induction}\\
&&\bm{F} = -\frac{c\lambda}{(\bar{\kappa}_{\rm ff}^{\rm R} + \kappa_{\rm es})\rho} \nabla E \label{eq:FLD_limiter}
\end{eqnarray}

The quantities $\rho$, $e$, and $\bm{v}$ are the density, internal energy, and velocity field of the gas. The pressure of the gas $p$ is related to the internal energy by $p = (\gamma - 1)e$, with adiabatic index $\gamma$ (here $\gamma = 5/3$ is assumed). The quantity $q$ is the pressure associated with a small artificial bulk viscosity. The quantity $\tilde{Q}_{\rm diss}$ represents the rate at which the internal energy must be changed for the total energy to be conserved
\citep[see Appendix A in][]{hir06}. The magnetic field is denoted by $\bm{B}$. 

The radiation field is described by the energy density $E$, flux $\bm{F}$, and pressure
tensor $\mathsf{P}$.  In the FLD approximation, the equations are closed with the relation
$\mathsf{P} = \mathsf{f}E$. The Eddington tensor $\mathsf{f}$ is a function of the
dimensionless opacity parameter $R \equiv |\nabla E|/[(\bar{\kappa}_{\rm ff}^{\rm R} +
\kappa_{\rm es})\rho E]$ and the flux limiter $\lambda$.  Here we adopt
$\lambda(R) = (\coth R - 1/R)/R$ (Levermore \& Pomraning 1981). A more complete
description of the FLD approximation is found in \cite{tur01}.

The gas and radiation exchange both momentum and energy via electron scattering and free-free emission/absorption.  For free-free absorption, the Planck-mean opacity and Rosseland-mean opacity are computed as $\bar{\kappa}_{\rm ff}^{\rm P} = 3.7\times 10^{53}(\rho^{9}/e^{7})^{1/2}$ cm$^{2}$/g and $\bar{\kappa}_{\rm ff}^{\rm R} = 1.0\times 10^{52}(\rho^{9}/e^{7})^{1/2}$ cm$^{2}$/g.  The electron scattering opacity is assumed to be $\kappa_{\rm es} = 0.33$ cm$^{2}$/g.  Energy exchange via Compton scattering is represented by the third term in the RHS of the energy equations, where $T$ and $T_{\rm rad} \equiv (E/a)^{4}$ are the gas and radiation temperatures, respectively \citep{bla03}.  The quantity $B$ in the
free-free energy exchange terms is defined as $caT^4/(4\pi)$.

To simulate the dynamics in a section of an orbiting disk, we adopt the shearing box
approximation, in which the sum of the Coriolis force, the tidal force, and the vertical
gravitational force, $\bm{f}_{\rm SB} \equiv -2\rho(\Omega\hat{\bm{z}})\times\bm{v} +
3\rho\Omega^{2}x\hat{\bm{x}} - \rho\Omega^{2}z\hat{\bm{z}}$, is added to the RHS of the
momentum equation.  Neglecting orbital curvature, we denote the radial coordinate, azimuthal
coordinate, and vertical coordinate by the Cartesian coordinates $x$, $y$, and $z$. 

We solved these equations using a modified version of the ZEUS code for radiation MHD
described in \cite{hir06}.  This version differs from the standard implementation of ZEUS
in two main respects.  First, strict energy conservation is achieved by capturing all
numerical dissipation of kinetic and magnetic energy, $\tilde{Q}_{\rm diss}$, and delivering
it to the gas's internal energy.  Second, the radiation diffusion equation is solved
in time-implicit fashion using a multi-grid algorithm.  Both of these alterations
are described in Appendix A of \cite{hir06}.

\subsection{Parameters}

Two parameters determine the problem in this context: the angular velocity $\Omega$
and the surface density $\Sigma$.  Using standard orbital mechanics and the
radiation-dominated limit of the $\alpha$ model, these quantities can be written
as functions of the central black hole mass $M$, the radius $r$, and the mass accretion
rate $\dot{M}$ for a guessed value of $\alpha$ (Krolik 1999):
\begin{equation}
\Omega = \sqrt{\frac{GM}{r^{3}}}
\label{eq:omega}
\end{equation}
\begin{equation}
\Sigma = \frac{8}{3\alpha\kappa_{\rm es}}\frac{\eta}{\dot{M}/\dot{M}_{E}}\left(\frac{r}{GM/c^{2}}\right)^{\frac{3}{2}}
\label{eq:sigma}
\end{equation}
As we have already explained, in a time-steady disk, the assumptions of radiative
support against vertical gravity and energy conservation can be used to relate
the disk vertical thickness $H$ and the radiated flux $F_0$ to $\dot M$ and $\Omega$:
\begin{equation}
F_{0} = \frac{3}{8\pi} \dot M \Omega^{2}
\label{eq:F0}
\end{equation}
\begin{equation}
H = \frac{3}{8\pi}\frac{\kappa_{\rm es}\dot M}{c}.
\label{eq:H}
\end{equation}
However, properly speaking, $H$ and $F_0$ (as well as $\dot M$) are driven by
local disk dynamics and are
in general time-dependent; thus, these relations should be regarded as guesses
that may or may not be confirmed.  For this reason, we use the $\alpha$ model
prediction for $H$ merely as an order-of-magnitude
estimator of the actual vertical thickness, and adopt it as our unit of length.

The specific parameters we chose for the simulations reported here are summarized in Table~1.
The efficiency $\eta$ is assumed to be 0.1.

\begin{deluxetable*}{llll}
\startdata
\tableline\tableline
Parameters & Definition      & Value                & Comment                    \\
\tableline
$M$        &$6.62\,M_{\sun}$ & $1.31\times10^{14}$g & Mass of central black hole \\
$r$        &$30\,(GM/c^{2})$ & $2.93\times10^{7}$cm & Radius                     \\
$\dot{M}$  &$0.1\,(L_{\rm E}/c^{2})/\eta$ & $7.52\times10^{17}$g s$^{-1}$ & Accretion rate \\
$\alpha$   &                 & 0.0125               & Stress/pressure ratio      \\
\tableline
$\Omega$   & eq.~\ref{eq:omega}          & $1.90\times10^{2}$ s$^{-1}$ & Orbital frequency          \\
$F_{0}$    & eq.~\ref{eq:F0}          & $4.62\times10^{21}$ erg cm$^{-2}$ s$^{-1}$ & Surface radiation flux     \\
$\Sigma$   & eq.~\ref{eq:sigma}          & $5.32\times10^{4}$ g cm$^{-2}$ & Surface density            \\
$H$        & eq.~\ref{eq:H}          & $1.46\times10^{6}$ cm & Disk scale height          \\
\tableline
\enddata
\end{deluxetable*}

\subsection{Grid and Boundary Conditions}

The size of the computational domain is $(L_{x},L_{y},L_{z}) =
(0.45H, 1.8H, 8.4H)$. It is divided into $(48,96,896)$ cells with a constant
cell size $\Delta x = \Delta z = \Delta y / 2$. The azimuthal boundaries are
periodic and the radial boundaries are shearing-periodic in the local shearing-box
appproximation \citep{hgb95}.  The vertical boundaries are (with one exception)
treated exactly as in \cite{hir06}: hydrodynamic quantities are given outflow
(free) boundary conditions, while the diffusive radiation flux and the
electric field do not change across the boundary.  We refer the interested
reader to the lengthy discussion found in that paper regarding the difficulties
of treating a subsonic boundary and the requirement for introducing a small
resistivity in the boundary cells.   The only difference between our treatment
and the one described there is that the artificial resistivity is extended
into the problem volume, tapering smoothly from the level in the ghost cells
to zero at 32 cells from the boundary \citep{kro07}.

\subsection{Initial Conditions}\label{sec:initial_condition}

The disk segment in the simulation box is initially assumed to be in both
dynamical and thermal equilibrium in the vertical direction, ignoring any
magnetic forces.  The key assumption underlying the structure of the initial
condition is the vertical profile of the dissipation rate; following \cite{hir06},
we assume that $Q_{\rm diss}$ is proportional to $\rho/\sqrt{\tau}$, where $\tau$
is the optical depth from a given point to the nearest surface.
The following equations are then solved to compute the vertical profiles of gas and radiation field for $1 \leqq \tau \leqq \tau_{0}$, where $\tau_{0}$ is the optical
depth to the midplane.  It is related to the surface density $\Sigma$ by $\tau_{0} \equiv (\Sigma/2)\kappa_{\rm es} + 1$; for our parameters, $\tau_0 = 8.79\times10^{3}$.
\begin{eqnarray}
\frac{d\rho}{d\tau} &=& \frac{\mu}{RT}\left(\frac{\Omega^{2}z}{\kappa_{\rm es}} - \frac{F}{c}\right) - \frac{3\rho F
}{4cE} \\
\frac{dz}{d\tau} &=& -\frac{1}{\kappa_{\rm es}\rho} \\
\frac{dE}{d\tau} &=& \frac{3F}{c}  \\
\frac{dF}{d\tau} &=& -\frac{Q_{\rm diss}}{\kappa_{\rm es}\rho}
\end{eqnarray}
Our procedure is to begin with the fourth equation, for which we impose the boundary conditions $F(\tau_{0}) = 0$ and $F(1) = F_{0}$; the result of solving this equation is the flux profile $F(\tau)$.  Then the radiation energy profile $E(\tau)$ is obtained immediately from the third equation. Next, the first two equations are solved numerically with boundary conditions $z(\tau_{0}) = 0$ and $\rho(1) = \rho_{\rm floor}$, to get the density profile $\rho(\tau)$ and the (inverse) opacity profile $z(\tau)$. The density floor $\rho_{\rm floor}$ is set to $10^{-5}\rho(\tau_{0})$. The gas temperature $T$ is assumed to be equal to the radiation temperature $(E/a)^{1/4}$. Then the midplane density and the height of the disk surface are obtained from $\rho(\tau_{0}) = 5.66\times10^{-2}$~g~cm$^{-2}$ and $z(1)/H = 1.36$.  Outside the photosphere ($z \geq z(1)$), we assume a uniform gas density $\rho_{\rm floor}$, a gas temperature $ T(z(1))$, and a radiation energy density $E(z(1))$. 

The initial velocity field is determined so that the inertial force is zero
($-2\rho(\Omega\hat{\bm{z}})\times\bm{v} + 3\rho\Omega^{2}x\hat{\bm{x}} = 0$), i.e.,
it is only the orbital shear $\bm{v} = (0, -3\Omega x/2, 0)$. The initial configuration
of the magnetic field is a twisted azimuthal flux tube (with net azimuthal flux)
placed at the center of the simulation box. The field strength in the tube is uniform
and the ratio of the poloidal field to the total field is 0.25 at maximum. The flux
tube cross section is an oval with radius in $x$ of $0.75(L_{x}/2)$ and radius
in $z$ of $0.80z(1)$.  The initial box-integrated plasma 
$\beta \equiv ((E/3) + p)/(\bm{B}^{2}/8\pi) = 11.5$, while in the midplane in the
initial state it is 40.
The corresponding maximum MRI wavelength in the midplane is resolved with 18.7 grid
zones.

The calculation is begun with a small random perturbation in the poloidal velocity. The maximum amplitude of each velocity component is $1\%$ of the local sound speed defined as 
$\sqrt{((4/3)(E/3) + \gamma p)/\rho}$.  
Two nearly identical simulations were run for just over 600 orbits each, the
only difference between them being the seed for the random initial poloidal
velocity perturbations.  
We refer to these two 
simulations as 1112a and 1126b, and we now discuss their properties.

\section{Simulation Results}


\subsection{Technical Consistency}

\begin{figure}
\plotone{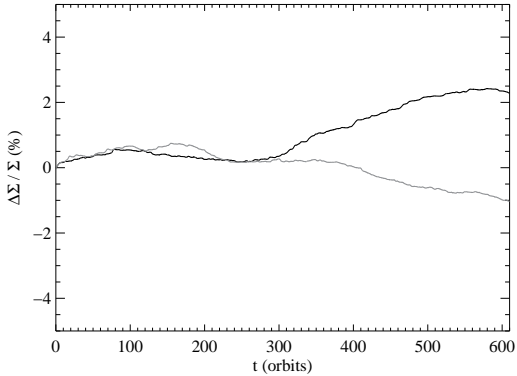}
\caption{Variation (in percent) of surface mass density as a function of time
$(\Sigma(t)-\Sigma(t=0))/\Sigma(t=0)$ in 1112a(black) and 1126b(gray).
\label{fig:surface_density_variation}}
\end{figure}

We are interested in the stability of disk segments in which the surface mass
density is constant.  To keep $\Sigma$ steady, the box height must be chosen
carefully because our boundary conditions permit outflows at the top and bottom of
the simulation box, but forbid inflows.  Consequently, mass is lost comparatively
readily in a shorter box, but mass can be added in a taller box due to more frequent
application of the code's density floor when inflows are shut off.  In simulations
1112a and 1126b, we set the box height at $8.4H$, which results in variations of the
surface mass density of less than
$2.5\%$ (Fig.\ref{fig:surface_density_variation}).

\begin{figure}
\plotone{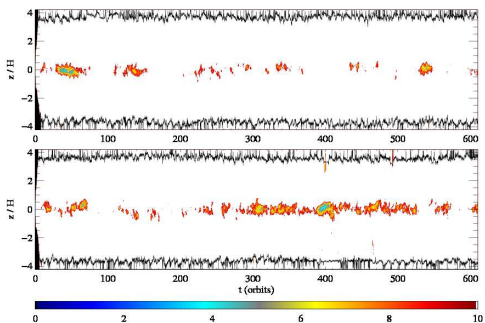}
\caption{Top panel: Number of cells per wave length of the most unstable mode of MRI in a horizontally averaged sense (color) and time variation of the photosphere location defined as the Eddington factor $f = 0.5$ (solid curves) in 1126b. Bottom panel: Same as the top panel except for 1112a.
\label{fig:MRIresolution_photosphere}}
\end{figure}

We must also ensure that the MRI is well-resolved throughout the box at all times.
The fastest growing mode has a vertical wavevector and wavelength
$\lambda_* = 6.49 / \Omega \sqrt{B_{z}^{2}/\rho}$.  Averaging $\lambda_*$ over
horizontal planes, we show the number of grid-cells per wavelength of this mode
in Fig.~\ref{fig:MRIresolution_photosphere}.  There are at least 8 cells per
wavelength (and generally a great many more) everywhere in the box except near
the midplane during a few brief epochs (e.g. $T\sim 400$ in 1112a).

The quality of our results also depends on keeping both the top and bottom photospheres
within the simulation domain at all times.  Fig.~\ref{fig:MRIresolution_photosphere}
also shows the locations of the photospheres in both simulations; we define
the photosphere by the place where the Eddington factor $f(\bar{R})=0.5$, for
$\bar{R}$ the horizontally-averaged dimensionless opacity parameter.  The photospheres
are comfortably within the computational domain throughout the duration of
both simulations.

\subsection{Thermal Balance}

\begin{figure}
\plotone{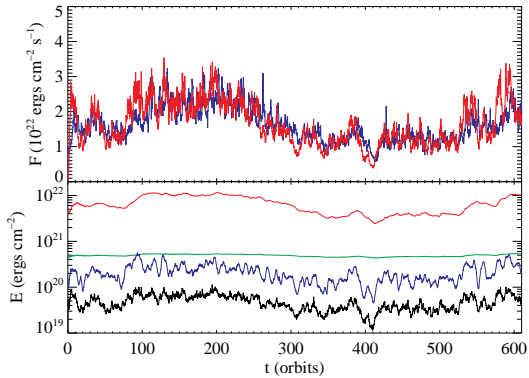}
\caption{Top panel:  Total volume integrated dissipation (red) and radiative
cooling (blue) rates as a function of time in simulation 1112a, plotted on
a linear vertical scale.  Bottom panel:   Total energy content in the box
as a function of time in simulation 1112a, in the form of radiation energy
(red), gas internal energy (green), magnetic energy (blue), and turbulent
kinetic energy (black).  (The turbulent kinetic energy is defined as the
kinetic energy associated with the fluid velocity measured relative to the
background shear flow.)  Note that the bottom panel is plotted on a logarithmic
vertical scale.
\label{fig:fluxesenergiesvst1112a}}
\end{figure}

\begin{figure}
\plotone{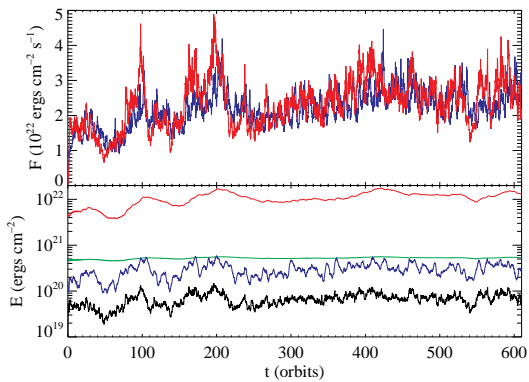}
\caption{Same as Fig. \ref{fig:fluxesenergiesvst1112a}, except for simulation
1126b.
\label{fig:fluxesenergiesvst1126b}}
\end{figure}

The thermal history of the two simulations is shown in Figs.
\ref{fig:fluxesenergiesvst1112a} and \ref{fig:fluxesenergiesvst1126b}.
The top panel in each figure shows the volume integrated dissipation rate
per unit surface area of the box (red) and the total radiative flux leaving
the top and bottom faces of the box (blue), both on a linear scale.  This
panel shows clearly that the box is able to maintain a close thermal
balance on timescales very short compared to the durations of these simulations.
The bottom panels show the total radiation, gas internal, magnetic, and turbulent
kinetic energies in the box on a logarithmic scale to emphasize better the
relative fluctuations in each of these quantities.  If we
define the unit of energy content as the gas thermal energy and consider
both simulations together, the radiation energy can range from $\simeq 5$--30,
the magnetic energy from $\simeq 0.15$--1, and the kinetic energy from
$\simeq 0.05$--0.25.  Adopting a nominal radiative efficiency of 0.1,
we find that the mean stress in simulation 1112a would drive
an accretion rate relative to Eddington of 0.17, while in 1126b it
would have been 0.23.  Thus, even when averaging over relatively
long timescales, the mean accretion rate for this surface density
and orbital frequency is still uncertain at the level of several
tens of percent.

A number of interesting results are apparent in these figures, but the one
that is most significant is the fact that an approximate thermal equilibrium
has been established in both simulations.  The thermal time as measured
by the radiation and gas internal energies divided by the total radiative
flux has short time scale fluctuations, but averaging over timescales
$\sim 50$~orbits, we find that in both simulations the instantaneous
cooling time can be anywhere from $\simeq 10$~orbits at times of relatively
low energy content to $\simeq 20$~orbits at times of especially high energy
in the box.  Averaged over the entire duration, the cooling time is 13~orbits
in 1112a, and slightly longer, 15.5~orbits, in 1126b.  Thus, both
simulations ran for $\simeq 40$ thermal times, and yet there is no evidence
of runaway heating and cooling.  \citet{sha76} predicted that thermal instability
should occur whenever the ratio of radiation to gas pressure exceeds 3/2,
but the radiation to gas energy ratios encountered here translate to pressure
ratios $\sim 2.5$--15, well above the predicted instability threshold.

Although it does appear that a crude long-term mean value for the energy
content can be estimated, the amplitude of fluctuations on timescales
a significant fraction of the simulation duration is still quite large.
We quantify these fluctuations by several measures.

First, we note that transients associated with the initial condition
are generally erased by $\simeq 100$ orbits into the simulation.
Discounting that initial period, we see that the typical fluctuation
amplitudes in the radiation, magnetic, and kinetic energy are similar,
all factors of several, but the fluctuation levels in the gas energy
are much smaller, only $6\%$~rms.  This means that the dissipated magnetic
and kinetic energies are readily converted to radiation energy.
On the other hand, although
there is considerable high-frequency power in the magnetic and kinetic
energy fluctuations, there is much less in the radiation energy.  That
this should be so is not surprising: the turbulence driving the magnetic
and kinetic energy densities has a characteristic timescale of order
an orbit, whereas the radiation energy changes on a thermal timescale,
an order of magnitude longer.

\begin{figure}
\plotone{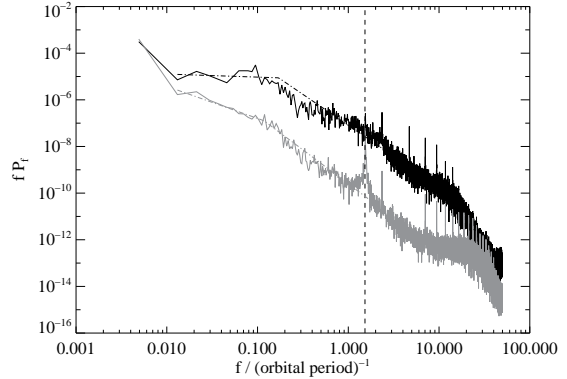}
\caption{Variability power spectral densities of volume-integrated magnetic
energy (black) and radiation energy (gray) in simulation 1112a.  Both
time-dependent energies were normalized by their time-averaged
values before computing the power spectral densities, and both power spectra
are plotted after having been multiplied by the frequency.  The vertical
dashed curve indicates the expected frequency of the longest wavelength
vertical acoustic wave.  Dot-dashed curves are broken power law fits
to the two power spectra in the frequency range 0.01 to 10.0, scaled
by inverse orbital periods.
\label{fig:erdemgpowerspec}}
\end{figure}

These remarks can be quantified by studying the power spectra of
the fluctuations (Fig.~\ref{fig:erdemgpowerspec} illustrates the
case of 1112a; 1126b is qualitatively similar).  The offset
between the magnetic and radiation Fourier power spectra is a measure of
the larger contribution from short timescale fluctuations in the time
series of magnetic energy.  Only at the very longest timescales does the
variance in the radiation energy become similar to that in the magnetic
energy (and also at very short timescales, where the fluctuation amplitude
is extremely small).  More quantitatively, both power spectra
can be approximately described by broken power-laws:
\begin{equation}
P(f) \simeq \cases{f^{-1.13,-3.65}; f_{\rm break} = 0.171 & magnetic\cr
                   f^{-2.38,-3.91}; f_{\rm break} = 0.118 & radiation\cr}.
\end{equation}
The first number in each exponent is the low-frequency slope; the second
number is the high-frequency slope.  In both cases, the break occurs at
$\simeq 5$--10 orbital periods, about half the thermal timescale.  At
high frequencies, both
power spectra decline quite steeply.  On the other hand, at low frequencies,
both power spectra are ``red" enough that the variance is dominated by
the low-frequency cut-off posed by the duration of the simulation.
This ``infrared divergence" is weak---almost logarithmic---for
the magnetic fluctuations, but is rather stronger for the radiation spectrum.
As we will argue in \S~4, the significant power at very long timescales
in the magnetic fluctuations is an important element in explaining
these disk segments' thermal stability.  As we will also argue, the
fact that the two power spectra approach one another at the lowest
frequencies is a symptom of the fact that MHD turbulence drives both,
so that they must approximately coincide on timescales longer than the
thermal equilibration time.

\begin{figure}
\plotone{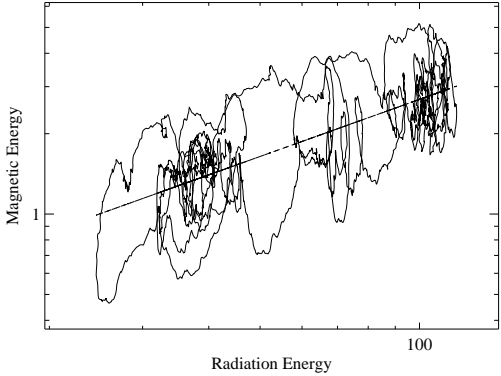}
\caption{Box-integrated magnetic energy and radiation energy compared
every 0.01~orbit from $T=100$ to $T=600$ in simulation 1112a.  Both energies
are in units of $10^{20}$~erg~cm$^{-3}$.  The
dashed line is a least squares fit in the logarithms of these two quantities.
\label{fig:emag_erad}}
\end{figure}

\begin{figure}
\plotone{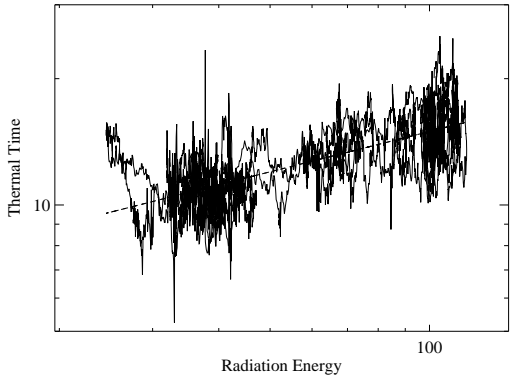}
\caption{Thermal time (in units of orbits) and radiation energy (in units of
$10^{20}$~erg~cm$^{-3}$) compared
every 0.01~orbit from $T=100$ to $T=600$ in simulation 1112a.  The
dashed line is a least squares fit in the logarithms of these two quantities.
\label{fig:tcool_erad}}
\end{figure}

The essence of the $\alpha$ model is the expectation that the stress should
be proportional to pressure.  We examine the validity of this expectation in
Fig.~\ref{fig:emag_erad}, but using the box-integrated magnetic energy
as a stand-in for the stress.  As this figure shows, the two do tend to
vary together, but with order unity fluctuations about the trend.  The
logarithmic slope of the least squares fit shown in the figure is 0.71,
indicating that the relation is shallower than linear.  For future reference
(this relationship will be called upon in \S~4), we also display (Fig.~\ref{fig:tcool_erad})
the analogous relationship between $t_{\rm cool}$ and the integrated radiation
energy.   The logarithmic slope of the fit in that case is 0.32.  It is worth
pointing out that the slopes of these relations are subject to some
chaotic variation: the magnetic--radiation energy correlation has slope 0.79
and the thermal time--radiation energy correlation has slope 0.44 in simulation
1126b.

\begin{figure}
\plotone{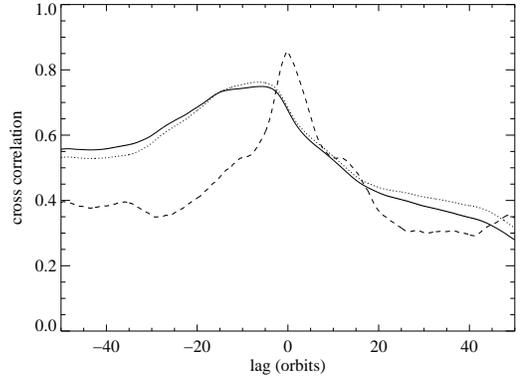}
\caption{Cross correlation with volume integrated magnetic energy of 
volume integrated radiation energy (solid), gas internal energy (dotted),
and turbulent kinetic energy (dashed) as a function of time lag for
simulation 1112a.  Negative
lag implies that the energy in question lags behind the magnetic energy.
\label{fig:energiescrosscorr}}
\end{figure}

Casual study of Figs.~\ref{fig:fluxesenergiesvst1112a},
\ref{fig:fluxesenergiesvst1126b}, and \ref{fig:emag_erad} suggests that,
after allowance for the
smoother variations in the radiation energy, all the contributions to
the energy content are well-correlated in time.  This is almost true.
A cross-correlation between them (Fig.~\ref{fig:energiescrosscorr})
reveals, however, that fluctuations in the magnetic energy {\it lead}
those in the radiation energy (and those in the gas energy)
by 5--15~orbits, roughly a thermal time.
In \S~4, we will show that this fact is a crucial clue for understanding the
boxes' thermal stability.  The turbulent kinetic energy, on the other
hand (and unsurprisingly) is closely coincident in time with the
turbulent magnetic energy.

We close this subsection by remarking on the quasi-periodic
oscillation (QPO) that appears in the power spectrum
of radiation energy at a frequency slightly higher than the orbital frequency.
We believe this to be a vertical acoustic oscillation.  Standing vertical
acoustic waves in a polytropic, Newtonian, Keplerian disk can occur at
angular frequencies given by
\begin{equation}
\omega=\Omega\left[{n_z(n_z+2n-1)\over2n}\right]^{1/2}
\end{equation}
where $n_z$ is an integer greater than or equal to two
and $n$ is the polytropic index \citep{kat05,bla06}.
The lowest frequency of this spectrum, $n_z=2$, fits the QPO exactly for
a radiation pressure supported configuration ($n=3$).  The eigenfunction
of this mode takes the form of a vertical breathing mode, with fluid
displacements expanding and contracting symmetrically about a node at the
midplane.  As we discuss in section 3.4 below, this mode plays an important
role in reconciling the actual vertical profiles of turbulent stress and
dissipation with the constraints of thermal and hydrostatic equilibrium.

\subsection{Stress}

In the original form of the $\alpha$-model, it is supposed that the
stress bears a fixed ratio to the total pressure.  However, it has
also been proposed \citep{sak81,taam84}, usually with a view toward quenching the
thermal instability, that the stress might instead scale more closely
with the gas pressure alone, or with the geometric mean of the gas
and radiation pressures.  We can test these hypotheses with our
simulation data.

\begin{figure}
\plotone{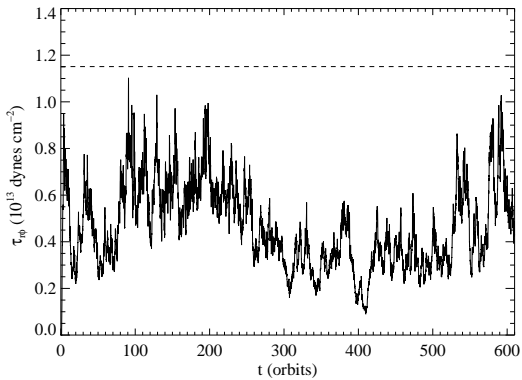}
\caption{Time dependence of the box-averaged $r\phi$ stress in simulation
1112a.  The horizontal dashed line indicates the $2c\Omega/3\kappa$ stress
expected from hydrostatic and radiative equilibrium.
\label{fig:taurphi1112a}}
\end{figure}

Fig.~\ref{fig:taurphi1112a} shows the time dependence of the 
box average of the $r\phi$ component of the Maxwell and turbulent
Reynolds stress in simulation 1112a.
The fact that it has considerable short timescale fluctuation power
immediately rules out the possibility that it should respond to either
the gas pressure or the radiation pressure in all respects because
their histories show only very weak short timescale fluctuations.
The large amplitude fluctuations in the stress over long timescales
similarly show that it cannot be driven, even in an averaged sense,
by the gas pressure, for the fluctuations in the gas pressure are
far too small.  That the volume-averaged stress is at all times
well below the critical stress
for pure radiation support demonstrates that, although radiation
pressure truly dominates in the central part of the disk, other
mechanisms (notably magnetic pressure, as we will show in \S~3.4) account
for support against vertical gravity in substantial portions of the
total simulation volume.

\begin{figure}
\plotone{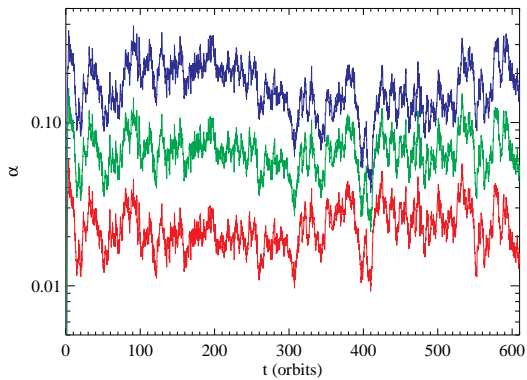}
\caption{Time dependence of the ratio $\alpha$ of the box-averaged
$r\phi$ stress to various box-averaged pressures in simulation
1112a.  From top to bottom, the stress is scaled with gas pressure,
the geometric mean of gas and radiation pressures, and with total (gas
plus radiation) pressure, respectively.
\label{fig:alphavst1112a}}
\end{figure}

To make this point very explicit, we show in Fig.~\ref{fig:alphavst1112a}
the time-dependence of the ratio of the box-integrated stress to the different
box-integrated candidate pressure quantities for simulation 1112a.
That is, these curves show the time-dependent Shakura-Sunyaev $\alpha$
parameter for various stress prescriptions.  The best would presumably
be the one that gives a value of $\alpha$ that varies least with time.
The least variation (taking the stress divided by the gas plus radiation pressure)
is a factor of 4, while the greatest variation (found when the stress is
measured in units of the gas pressure) is a factor of 10.  In {\it no}
case, therefore, does the stress follow exactly the fluctuations in any of
these pressure measures.

It is worth remarking in this context that the mean value of the ratio
of integrated stress to integrated total pressure was 0.023 in 1112a
and 0.019 in 1126b.  In our previous simulations, this number was
$\simeq 0.03$ (for the $p_r \sim p_g$ simulation reported in \citet{kro07})
and 0.016 (for the $p_r/p_g \simeq 0.2$ simulation of \citet{hir06}).
Thus, we see little trend in this ratio as a function of the ratio of
radiation to gas pressure.  In addition, this series of simulations
is, somewhat inadvertently, a crude test of numerical convergence.
Due both to the requirements of the simulations and to the increase in
available computational power over time, the resolutions employed
have improved steadily from one to the next.  Measured in terms of
our estimated scale-height, the cell-size $\Delta z$ changed from
$0.0625H$ in the gas-dominated simulation to $0.0234H$ in the
$p_r \sim p_g$ simulation to $0.0094H$ in the simulations presented
here.  The rough constancy of the stress/pressure ratio across these
simulations is of some interest in view of reports that {\it unstratified}
shearing box simulations show a declining ratio of stress to pressure as
resolution improves \citep{fro07}.  It should be pointed out, however,
that these unstratified simulations had zero net magnetic flux, whereas
our stratified simulations have a net azimuthal flux, albeit one that
is not preserved by the boundary conditions and
fluctuates in direction over the course of the simulations.

\subsection{Vertical Energy Transport and Hydrostatic Balance}

\begin{figure}
\plotone{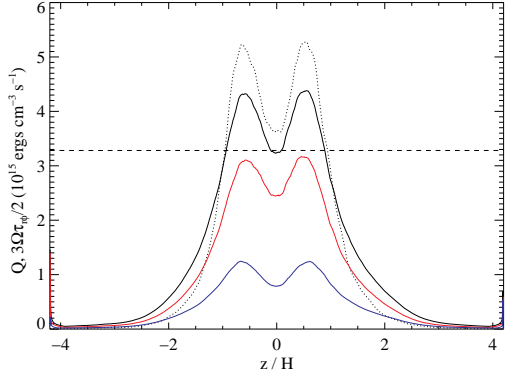}
\caption{Horizontally and time-averaged total dissipation rate $Q$
(black) and stress times rate of strain (dotted) in simulation 1112a.
The total dissipation rate is the sum of magnetic energy dissipation
(red) and kinetic energy dissipation (blue). 
The horizontal dashed line marks the characteristic dissipation rate
$Q_*=c\Omega^2/\kappa$.
\label{fig:dissprofile}}
\end{figure}

As first pointed out by \citet{sha76}, in any radiation-dominated geometrically-thin
disk, there is a characteristic emissivity per unit volume:
$j_* = c\Omega^2/\kappa$.  Its existence follows immediately from attributing
hydrostatic balance to radiation force.  When that condition is met (and
the opacity is purely electron scattering),
\begin{equation}
\kappa_{\rm es} {F_{z}}/c = z\Omega^2;
\end{equation}
differentiating both sides with respect to height $z$ gives
\begin{equation}
{\kappa_{\rm es} \over c}{d{F_{z}} \over dz} \equiv {\kappa_{\rm es} \over c}j = \Omega^2.
\end{equation}
If all the dissipation is transformed locally into
photon energy, there is a corresponding characteristic dissipation rate $Q_* = j_*$.
We find (Fig.~\ref{fig:dissprofile}) that the dissipation rate is comparable to
$Q_*$ through most of the disk body, but is typically somewhat greater at
$|z| \simeq 0.5H$ (by $\simeq 30\%$) and rather less in the upper disk.
At all altitudes it is dominated by magnetic losses.

\begin{figure}
\plotone{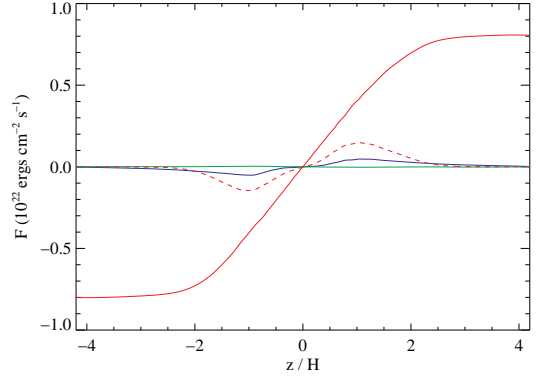}
\caption{Horizontally and time-averaged diffusive radiation energy flux $F_{z}$
(solid red), advected radiation energy flux $Ev_{z}$ (dashed red), Poynting
flux $(\bm{E}\times\bm{B})_{z}$ (blue), and advected gas energy flux $(e+\rho\bm{v}^{2}/2)v_{z}$ (green) in simulation 1112a.
\label{fig:fluxesprofile}}
\end{figure}

\begin{figure}
\plotone{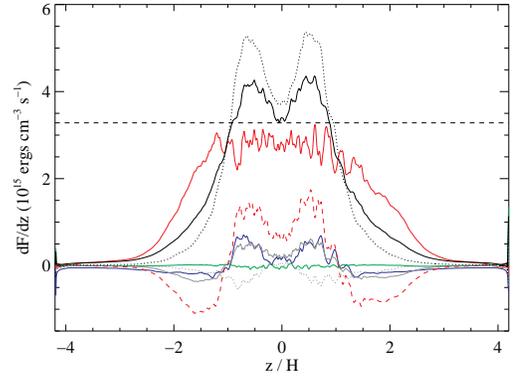}
\caption{Horizontally and time-averaged vertical derivatives of diffusive
radiation energy flux $F_{z}$ (solid red), advected radiation energy flux
$Ev_{z}$ (dashed red), Poynting flux $(\bm{E}\times\bm{B})_{z}$ (blue),
and advected gas energy flux $(e+\rho\bm{v}^{2}/2)v_{z}$ (green)
in simulation 1112a.  In addition, the grey curve shows the profile of
$\mathsf{P}:\nabla\bm{v}-\chi\rho\bm{F}\cdot\bm{v}/c$, which is equivalent
to the divergence of the flux $Ev_{z}/3$ of work done by radiation pressure
when the medium is optically thick.  The sum of the diffusive and advective
radiation energy flux derivatives is shown as the solid black line, and
matches well the horizontally and time-averaged dissipation profile $Q$ (solid
black line in Fig.~\ref{fig:dissprofile}).  The sum of all the flux derivative
profiles is shown as the dotted black line, and matches well the horizontally
and time-averaged profile of stress times rate of strain (dotted black
line in Fig.~\ref{fig:dissprofile}).
For reference, the horizontal dashed line marks the
characteristic dissipation rate $Q_*=c\Omega^2/\kappa$.
\label{fig:divfluxesprofile}}
\end{figure}

If hydrostatic balance obtains (as we will show shortly, departures are almost
always small), where $Q > Q_*$ over a significant range in height, it must be
that $j < Q$ because otherwise
the flux would exceed the value whose force balances gravity.  Consequently,
not all the energy liberated by dissipation goes directly into radiation
flux.   As shown in Fig.~\ref{fig:fluxesprofile}, for $|z|\lesssim2H$,
a noticeable (but minority: $\sim 10\%$) fraction of the vertical energy flux is
in the form of photons advected upward {\it with} the matter ($Ev_{z}$), rather than
diffusing through it ($F_{z}$); because this component does not move through the
matter, it exerts no force.  Like the total radiation content in the box,
the advective energy flux also oscillates with the frequency of the fundamental
vertical breathing mode.  If this oscillation behaved with exact symmetry in its
outward and inward moving phases, it would carry no net energy.  However,
at the outer surface
of the region of significant advection ($|z| \simeq 2H$), we see a divergence
of the diffusive flux that exceeds the local dissipation rate
(Fig.~\ref{fig:divfluxesprofile}); that is,
radiation energy diffuses out of the upwelling matter at the top of its
oscillatory range.  In this respect, the radiative advection seen here
resembles convection, although the vertical gas motions are acoustic in
nature and not due to buoyancy.  In addition, a smaller
amount of energy is carried upward by work done by radiation pressure forces
and by Poynting flux $(\bm{E}\times\bm{B})_{z}$,
which also oscillates with the frequency of the breathing mode.  Clearly, MHD
flux-freezing causes magnetic energy to be advected with the gas along with
radiation energy in the breathing mode.
The energy transported by radiation advection, work done by radiation
pressure,  and Poynting flux is deposited (or dissipated) around
$H\lesssim|z|\lesssim2H$.
Above this point, the energy flux is overwhelmingly due to radiation diffusion.
Energy transport by gas advection $(e+\rho\bm{v}^{2}/2)v_{z}$ and gas pressure
work is negligible throughout the simulation domain.

Fig.~\ref{fig:dissprofile} also shows the vertical profile of stress
times rate of strain, $-\tau_{r\phi}d\Omega/d\ln r$.  This profile is
significantly more concentrated near the midplane
and at $z = \pm 0.5H$ than the dissipation
profile, indicating that the work done by the turbulent stresses is, in
the time average, partly dissipated locally and partly converted into
mechanical energy of gas motion, which is then transported vertically
primarily by Poynting flux and radiation pressure work associated with
the breathing mode and dissipated farther out.  This energy is
deposited well outside the peak region of stress, and subsequently
carried further outward by radiative diffusion.  This supplemental
energy transport mechanism contributes to
$dF_{z}/dz$, helping to maintain radiation-aided
support against gravity at altitudes higher than one might expect
if dissipation correlated perfectly with local stress.

The net result of these processes is illustrated most clearly in
Fig.~\ref{fig:divfluxesprofile}. As this figure shows, despite the
peaks in $Q$ that reach well above $Q_*$, the
divergence of the radiation flux $F_{z}$ is kept just under $j_*$ across
the entire central body of the disk, from $z \simeq -1.5H$ to $z \simeq +1.5H$.
It is the radiation advection flux $Ev_{z}$ that compensates for
the excess dissipated energy $(Q-Q_{*})$ and carries it away. Thus, thermal
balance is locally maintained in the entire region, as evidenced by the fact
that the divergence of the summed radiation and gas energy fluxes
($F_{z} + Ev_{z}$) matches the dissipation rate $Q$
very well.  At the same time, Poynting flux $(\bm{E}\times\bm{B})_{z}$ and
the flux of work done by radiation pressure forces transports non-dissipated
excess mechanical energy away from the midplane to be dissipated further out.
The sum of the divergences of all these energy fluxes (together with advection
of gas energy and work done by gas pressure, which are both small) compensates
exactly for the work done on the fluid by turbulent stresses, thereby
establishing global energy conservation.

\begin{figure}
\plotone{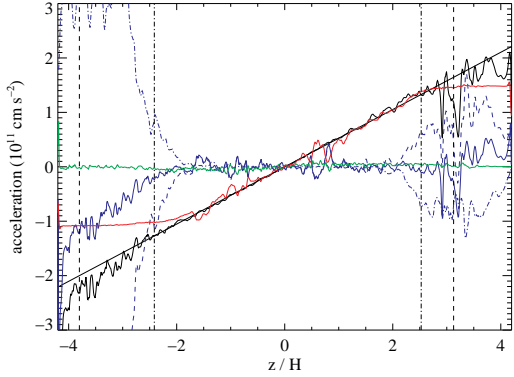}
\caption{Horizontally averaged vertical accelerations at the 200 orbits
epoch in simulation 1112a.  The different curves show accelerations from
radiation pressure (red), gas pressure (green), and magnetic forces (blue).
The total of these accelerations is the black curve, and is comparable to
the straight black line indicating the gravitational acceleration.  The
dashed blue curve is the contribution to the magnetic acceleration from
magnetic pressure gradients, while the dot-dashed blue curve indicates
the contribution from magnetic tension.  The vertical dashed and dot-dashed
lines indicate the locations of the scattering and thermalization photospheres
of the horizontally-averaged structure, respectively.
\label{fig:vertaccelerations}}
\end{figure}

Consistent with the rough match between $Q$ and $Q_*$, most of the vertical
support, particularly in the center of the disk, is due to radiation forces
(Fig.~\ref{fig:vertaccelerations}) and hydrostatic balance is maintained
quite closely.  In fact, departures from time-averaged hydrostatic balance
are smaller, in fractional terms, than any local deviation (again time-averaged)
between $Q$ and $Q_*$.  In general terms, this occurs because the timescale
for dynamical equilibration ($\sim 1/\Omega$) is the shortest timescale
in the system.  In detail, two different effects account for the fact that
deviations from hydrostatic balance are so small.  As just discussed,
in the disk body, where $Q > Q_*$, some of the vertical energy transport
is carried by other mechanisms, notably radiation advection
(Fig.~\ref{fig:fluxesprofile}).  At higher
altitudes, vertical support depends more on the gradient of magnetic pressure
than on the gradient of radiation pressure.

   The structure that is established by these several forces is very
similar qualitatively to the structure seen in all previous vertically-stratified
shearing-box simulations with genuine thermodynamics: a central region ($|z|<3H$)
supported by a combination of gas and radiation pressure lying within
an outer region dominated by magnetic forces.  Within the magnetically-dominated
region, the outward force due to the magnetic pressure gradient is
substantially cancelled (in the mean) by the inward force due to magnetic
tension (Fig.~\ref{fig:vertaccelerations}).

\begin{figure}
\plotone{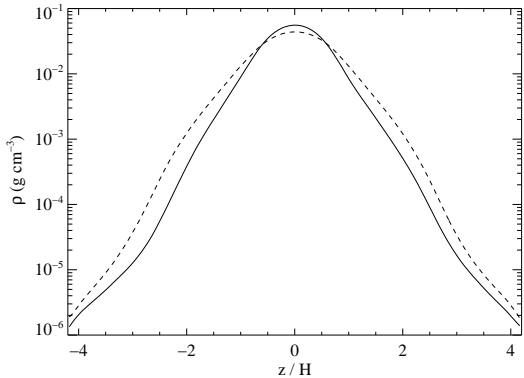}
\caption{Horizontally and time-averaged vertical density profile in the
1112a (solid) and 1126b (dashed) simulations.
\label{fig:densitytavg}}
\end{figure}

\begin{figure}
\plotone{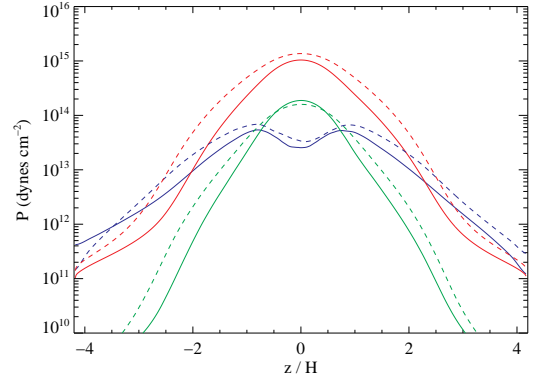}
\caption{Horizontally and time-averaged vertical profiles of radiation
pressure (red), gas pressure (green) and magnetic pressure (blue) in the
1112a (solid) and 1126b (dashed) simulations.
\label{fig:pressurestavg}}
\end{figure}

   The resulting profiles of density and pressure are roughly exponential
(Figs.~\ref{fig:densitytavg} and \ref{fig:pressurestavg}).  An exponential
description is particularly good for the density, with time-averaged
scale-height $\simeq 0.8H$.  However, we caution that the instantaneous
scale-height can vary by order unity as the radiation energy varies up
and down.  It is also appropriate to remark at this point that, contrary
to the simple assumption made in classical $\alpha$-models, the density
is very far from constant because the dissipation per unit mass is also
very far from constant.  As we have found in earlier vertically-stratified
studies with smaller ratios of radiation to gas pressure, the dissipation
rate per unit mass increases with decreasing mass column density to the
nearest outer surface.

   The pressures, while still behaving roughly as exponentials in height,
have additional structure.   The radiation pressure declines more steeply
within $|z|< 3H$ than without because the density drops so steeply outward.
From the photosphere outward, the radiation pressure changes
only slowly with height because that is the free-streaming limit in plane-parallel
geometry.  The magnetic pressure profile, on the other hand, is roughly flat-topped
in the central disk, with a pair of weak local maxima at $|z| \simeq H$.  In
that central region, the plasma $\beta \simeq 50$, but it drops rapidly
outward, falling below unity at $|z| \simeq 2.5H$ where the magnetic
pressure begins to dominate the total.

\begin{figure}
\plotone{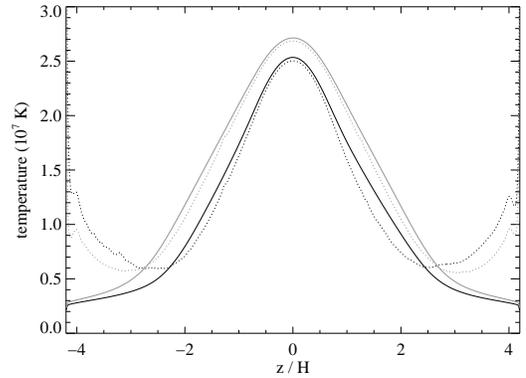}
\caption{Horizontally and time-averaged gas temperature (solid) and radiation
temperature (dotted) in 1112a (black) and 1126b (gray).
\label{fig:temperatures_0-610}}
\end{figure}

\begin{figure}
\plotone{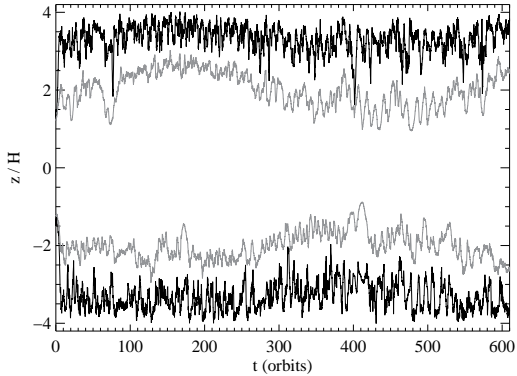}
\caption{Horizontally-averaged height of the scattering (black) and
thermalization (gray) photospheres as functions of time in 1112a.
\label{fig:photospheres}}
\end{figure}

Fig.~\ref{fig:temperatures_0-610} shows horizontally and time-averaged gas
temperature $T$ and radiation temperature $T_{\rm rad} \equiv (E/a)^{1/4}$,
while Fig.~\ref{fig:photospheres} shows the height of
the scattering and thermalization photospheres as functions of time for
simulation 1112a.  These photospheres are defined
as the surfaces where the horizontally averaged scattering optical depth
and geometric mean of the scattering and Planck mean absorption optical
depths, respectively, equal unity.  The time-averaged height of the
thermalization photosphere in simulation 1112a was $2.0H$, in approximate
agreement with the height
at which the time-averaged gas and radiation temperatures begin to separate
in Fig.~\ref{fig:temperatures_0-610}.  The scattering photosphere is of
course further out, with a time-averaged height of $3.3H$.

\begin{figure}
\plotone{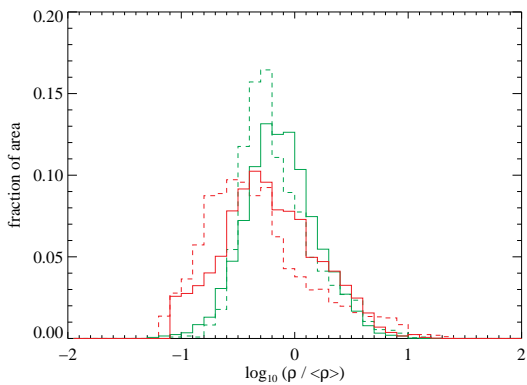}
\caption{Histogram of the ratio of density to horizontally-averaged density
at the thermalization (green) and scattering (red) photospheres of the
horizontally averaged structure at epoch 200 in simulation 1112a.
The upper and lower photospheres are indicated by solid and dashed histograms,
respectively.  The fluid motions are highly compressible in this region due to
the large ambient magnetic forces.
\label{fig:rhofluct}}
\end{figure}

Fig.~\ref{fig:rhofluct} shows that very large density fluctuations are always
present at both the thermalization and scattering photospheres of the
horizontally averaged structure.  Similarly large density
fluctuations in the photospheres have been seen in all our previous simulations
(see Fig. 16 in \citealt{bla07}), and are due to the fluctuating
magnetic forces that dominate all other forces in these regions.

\begin{figure}
\plotone{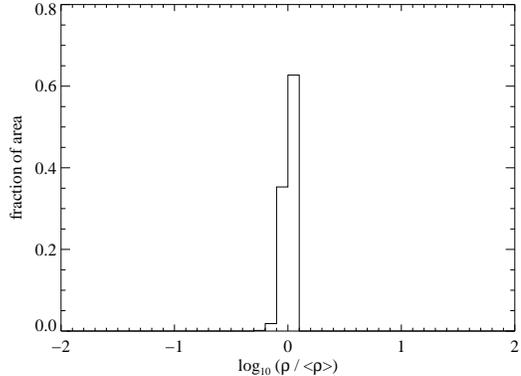}
\caption{Histogram of the ratio of density to horizontally-averaged density
at the midplane for simulation 1112a at epoch 200.  The fluid motions
are highly incompressible in this region.
\label{fig:rhofluctmidplane}}
\end{figure}

We close this section by remarking that we have confirmed many of the results
seen in the radiation pressure dominated simulation of \citet{tur04}, despite
the fact that that simulation was not fully energy conserving, nor did it
include the photospheres within the simulation domain.  In particular,
he also found no thermal instability and a vertical density profile that was
highly concentrated toward the midplane.  As shown in
Fig.~\ref{fig:rhofluctmidplane}, we also find
similarly small density fluctuations near the midplane, of at most a factor
$\sim2$ between maximum and minimum. The magnetic pressure profile in
his simulation had a double-peaked structure that was lower than the gas
pressure at the midplane, but exceeded the gas pressure further out, also
in agreement with what we have found.  However, his
magnetic pressure nowhere exceeded the radiation pressure, in contrast to
the outermost layers in our simulation.  Moreover, he found that
radiative diffusion carried only two-thirds of the outward energy flux, with
radiation advection being the most important secondary coolant.  These
differences can perhaps be ascribed to the fact that his simulation did not
include the photospheres.  If we neglected the data from our simulations
outside $|z|=2H$, we would see results qualitatively similar to his.

The only respect in which we really disagree is that \citet{tur04} also claimed
that radiation damping \citep{ago98} contributed $29\%$ of the heating in the
midplane regions of his simulation.  However,
we have demonstrated a complete accounting of the thermal energy
balance without calculating the contribution from this process at all.
Turner computed the radiation damping rate by integrating
$-P\nabla \cdot\bm{v}$ over the midplane region of his box, with $P=E/3$ where
radiation dominates and is isotropic.  This is a reasonable method in
a nonstratified shearing box \citep{tur02} because the volume is fixed:
consequently, the {\it net} compressive work indicates non-adiabatic
behavior.  However, in a stratified shearing box, $P\nabla\cdot\bm{v}$ is
really a piece of the divergence of the flux of work done by radiation
pressure, $\nabla\cdot(P\bm{v})$, and we have shown that this plays a
contributing role to transporting excess mechanical energy out of the
midplane by the breathing mode to be dissipated further out.  Moreover,
even if we follow Turner in using $-P\nabla\cdot\bm{v}$ to provide an 
estimate of the radiation damping near the midplane,
we then find that this can be at most $1.3\%$
and $0.7\%$ percent of the dissipation in the midplane region in 1112a and
1126b, respectively.  It is conceivable that one of two limitations in
Turner's code may be responsible for this quantitative difference between his
simulation and ours: the lack of a photosphere and the
inability to capture grid scale losses of kinetic energy.  Without performing
additional simulations that separately reproduce each of these possibilities,
we cannot be certain as to the exact origin of the discrepancy.
Nevertheless, we suspect that lack of true energy conservation in Turner's
simulation is the primary cause.

\section{Interpretation}

     In the traditional $\alpha$-model approach, whose logical underpinning
is dimensional analysis, the stress, and therefore the magnetic field energy
density, is scaled to the pressure.  Considered in a logically rigorous
fashion, dimensional analysis suggests that two quantities with the same
units are comparable
in magnitude, but does not determine a causal relation between the two:
that is, it does not tell us which quantity determines the other.  Nonetheless,
if it is easier to estimate one than the other, it is easy to slip into
the thought that the one harder to estimate follows the other, not just
in scale, but in time.  In this case, the $\alpha$-model has often been
imagined to suggest that the pressure controls the magnetic field.

     However, a crosscorrelation of the magnetic field energy against the
radiation energy in this simulation shows that fluctuations in the magnetic
field {\it lead} those in the radiation.  If so, the dominant sense of causality
must be the other way around---the magnetic field controls the radiation,
not vice versa.  In fact, that is the direction suggested by the dynamics:
The MRI causes field fluctuations to grow, deriving
the needed energy from orbital shear; nonlinear mode-mode interactions
move power from long lengthscales to short; short wavelength dissipative
effects transfer energy from the magnetic field to heat in the plasma;
finally, electrons in the plasma radiate photons.  In other words,
intrinsic fluctuations in the magnetic turbulence can create a pressure-stress
correlation even when there is no direct influence of the pressure on
the saturation level of the magnetic field.  All that is required is
that their amplitude on timescales long compared to a thermal equilibration
time are large enough, and these are already known to occur even in
gas-dominated cases in which the pressure varies very little \citep{hir06}.
When that is the case, the increased dissipation
rate due to the stronger field leads to higher pressure, but not
necessarily to any subsequent increase in the stress.  Thus, a
key assumption of the argument for thermal instability in the
radiation-dominated regime is undermined.

      Using an approach reminiscent of the long-wavelength limit
of the formalism developed in \citet{pir78}, the qualitative picture
outlined above can be modelled by the following pair of equations:
\begin{eqnarray}
\frac{dE_B}{dt} &=&  R(t){E_{B0}\over t_{\rm growth}}
                 \left(\frac{E_R}{E_{R0}}\right)^n
        - {E_B \over t_{\rm diss}}\\
\frac{dE_R}{dt} &=& {E_B \over t_{\rm diss}} - {E_R \over t_{\rm cool}},
\end{eqnarray}
where $E_{B,R}$ are the box-integrated magnetic and radiation energies,
$E_{B0,R0}$ are their equilibrium values, and $R(t)$ is a random function with
expectation value unity describing the intrinsic fluctuations in the scaling of
magnetic energy density with pressure.  We also suppose that the equilibrium
magnetic energy density grows with pressure (here the radiation energy) with
logarithmic derivative $n$.  Thus, we allow for feedback in the sense that the
pressure might be able to influence the stress.  Lastly,
$t_{\rm growth}$ is the timescale for the magnetic energy to adjust toward
its expectation value, $t_{\rm diss}$ is the timescale for magnetic
dissipation, and $t_{\rm cool}$ is the radiation cooling time.  The magnetic
field energy matches the equilibrium value when
$t_{\rm growth}=t_{\rm diss}$, so these two timescales may be equated.
Non-dimensionalizing the energies in terms of their equilibrium values
and the time in terms of $t_{\rm diss}$, these equations become
\begin{eqnarray}
\frac{d{\cal E}_B}{d\tau} &=& R(t){\cal E}_R^n  - {\cal E}_B \\
\frac{d{\cal E}_R}{d\tau} &=& {\cal E}_B\frac{E_{B0}}{E_{R0}} 
   - \frac{t_{\rm diss}}{t_{\rm cool}}{\cal E}_R .
\end{eqnarray}
The timescale ratio $t_{\rm diss}/t_{\rm cool}$ may also depend on pressure;
we designate its logarithmic derivative with respect to pressure by $-s$
(the minus sign makes its definition consistent with the notation in
\citet{kro07}).

That assumption puts the model equations in the form:
\begin{eqnarray}
\frac{d{\cal E}_B}{d\tau} &=& R(t){\cal E}_R^n - {\cal E}_B \\
\frac{d{\cal E}_R}{d\tau} &=& {\cal E}_B\frac{E_{B0}}{E_{R0}}
  - \frac{t_{\rm diss,0}}{t_{\rm cool,0}}{\cal E}_R^{1-s} .
\end{eqnarray}
Once again we can use the requirement that equilibrium values be consistent
with zero time-dependence to constrain these parameters: this time we conclude
that $t_{\rm diss,0}/t_{\rm cool,0} = E_{B0}/E_{R0}$, leaving
\begin{eqnarray}
\frac{d{\cal E}_B}{d\tau} &=& R(t){\cal E}_R^n - {\cal E}_B \\
\frac{d{\cal E}_R}{d\tau} &=& \frac{E_{B0}}{E_{R0}}\left[{\cal E}_B
       -{\cal E}_R^{1-s}\right] .
\end{eqnarray}
In other words, the timescale for the radiation energy to equilibrate
is $\sim E_{R0}/E_{B0}$ times the timescale on which the magnetic energy
can fluctuate; the magnetic energy must be dissipated many times over
in order to produce a response in the radiation energy because $E_{R0}/E_{B0}$
is so large.  For context, we recall that in the simulations described
here, the mean $E_R/E_B \simeq 30$ while the mean thermal timescale was
$\simeq 15$~orbits.  If this toy-model describes the simulations,
i.e., if heating in the simulations is primarily due to magnetic
dissipation and cooling is due to radiation losses, these numbers
would suggest that in the simulations $t_{\rm diss}$ should be
defined as the ratio of box-integrated magnetic energy to magnetic
dissipation rate, and its typical value should be
$\simeq 0.5$~orbits.  In fact, consistent with
this description of energy balance, the mean value of $t_{\rm diss}$
defined in this way was 0.53~orbits in simulation 1112a, and 0.55~orbits
in simulation 1126b.

\begin{figure}
\plotone{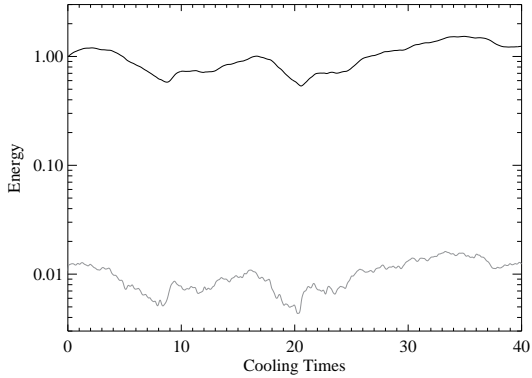}
\caption{Radiation energy (black curve) and magnetic energy (gray curve)
as functions of time in units of $t_{\rm cool}$, as computed on the basis
of the toy-model described in the text.
\label{fig:energyhist}}
\end{figure}

In Fig.~\ref{fig:energyhist}, we show a sample solution of these
equations, in this case with $n=s=0$, $E_{B0}/E_{R0} = 0.033$, and initial
conditions ${\cal E}_R = {\cal E}_B = 1$.  Note that setting $n=0$ means
that there is {\it no} explicit dependence of magnetic amplification on
the radiation pressure.  The random function $R(t)$ was
constructed by requiring its power spectrum to match approximately the power
spectrum of magnetic energy fluctuations seen in simulation 1112a (i.e.,
$\propto f^{-1.13}$ at frequencies 
$f \leq 2.6/t_{\rm cool}$ and $\propto f^{-3.65}$ for $f > 2.6/t_{\rm cool}$), 
but choosing the phases
of its Fourier amplitudes randomly with uniform probability distribution
in the interval $[0,2\pi)$.  To make the comparision as direct as possible,
the duration of the integration was set to $40t_{\rm cool}$.  Just as in
the simulations, the radiation energy fluctuates by factors of several
over this timespan, but shows no clear trend.

\begin{figure}
\plotone{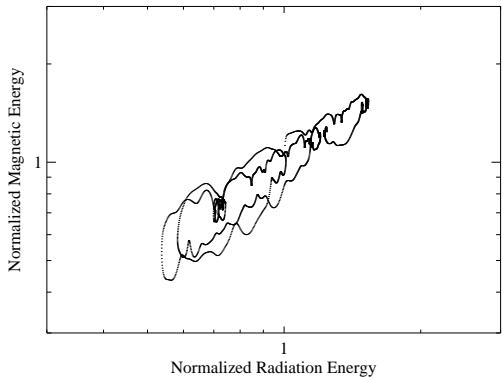}
\caption{Correlation plot between ${\cal E}_R$ and ${\cal E}_B$.
\label{fig:magradcorr}}
\end{figure}

Guided by a traditional view of the $\alpha$ model, one might object that
the absence of thermal instability here is due to the lack of any explicit
dependence of the magnetic energy on the radiation energy.  However, closer
study of the differential equations suggests that if $t_{\rm cool}$
is short compared to the timescale of the large-amplitude turbulence
fluctuations, approximate radiative equilibrium is enforced on the
shorter timescale.  It would then follow that ${\cal E}_B \propto
{\cal E}_R^{1-s}$ {\it independent} of $n$.  This is exactly what we see,
for these parameters and all other combinations of $n$ and $s$ we have tried,
subject only to the constraint that $n < 1 - s$ (as discussed in \citet{kro07},
genuine instability can be found when this limit is violated).  The flat power
spectrum at frequencies below $1/t_{\rm cool}$ guarantees that there is
substantial fluctuation power on timescales longer than the radiative
equilibration timescale, and the result is as predicted, as shown by
Fig.~\ref{fig:magradcorr}.  A linear least-squares fit in the logarithms
of these two quantities then yields a best-fit relation in which
$d\log{\cal E}_B/d\log {\cal E}_R \simeq 1-s$; in the example shown,
in which $s=0$, the best-fit logarithmic derivative was 1.01.
Thus, magnetic energy and pressure {\it do} scale together as suggested
by the dimensional analysis underlying the $\alpha$ model, but it is
{\it not} necessarily because the pressure
directly forces the magnetic energy; it is instead the other way around:
the pressure is the result of magnetic dissipative heating regulated by
photon losses.

We now see the data shown in Fig.~\ref{fig:emag_erad} in a different
light.  Although it is true that, compared at the same time,
$E_B \propto E_R^{0.71}$ in that simulation,
this correlation is created by the dependence of the radiation energy on the
magnetic energy, and not by the radiation pressure regulating the magnetic
field strength.  Moreover, if our simple model correctly describes the
thermal balance, the slope of this correlation should be $1-s$, so that
$t_{\rm cool}/t_{\rm diss} \propto E_R^{0.29}$.  We have already
seen that $t_{\rm cool} \propto E_R^{0.32}$; a logarithmic least-squares
linear fit to the data of 1112a gives $t_{\rm diss} \propto E_R^{0.02}$,
supporting the toy-model.

An important corollary of these results is that the intrinsic
dependence of magnetic energy on radiation pressure parameterized by $n$
{\it cannot} be inferred from the slope of the correlation between
magnetic energy and radiation energy; this slope is entirely independent of
it.  The most that one can do is to place an upper bound on how strongly
pressure influences magnetic saturation by using the observed thermal
stability to argue that $n < 1 -s$.

To close this section, we make the qualitative remark that the sensitivity
of the cooling time to pressure may influence the magnitude of fluctuations
in the energy content.  As argued in \citet{kro07}, provided
$t_{\rm diss}$ is not too strong a function of box energy content, $s$
must be relatively large in magnitude and negative when gas pressure
dominates radiation pressure because an increasing ratio of radiation
to gas pressure makes diffusive cooling much more rapid.  If so, the
cooling time rapidly shortens when the radiation energy increases,
placing a tight cap on fluctuations in the energy content of the box.
When the radiation to gas pressure ratio is order unity or greater,
the scaling of $t_{\rm cool}$ with total energy should be rather
slower (a prediction consistent with our results if $t_{\rm diss}$ is
likewise not a strong function of $E_R$).  In this case, the cooling
time is comparatively insensitive to energy content and exercises
looser control on energy fluctuations, permitting the turbulence
fluctuations to drive larger amplitude fluctuations in the total
pressure.

\section{Conclusions and Summary}

      In this paper we have presented the results of two simulations
that each followed the evolution of a vertically-stratified shearing box
with the same surface density and orbital frequency for $\sim 600$~orbits,
or $\sim 40$~thermal times.   The initial conditions of the two simulations
differed only in being given different realizations of the small amplitude
noise that was imposed on their otherwise smooth initial state.  Because
the magneto-rotational instability generates MHD turbulence, these systems
are chaotic; differing small amplitude noise therefore leads to order
unity contrasts in their subsequent evolution.  The two simulations should
thus be viewed as two different realizations of the many evolutions possible
for these parameters.  Their qualitative aspects are, nonetheless, similar.

For example, just as we have found in previous simulations studying
shearing boxes with $p_r \simeq 0.2p_g$ \citep{hir06} and $p_r \sim p_g$
\citep{kro07}, most of the disk mass is found near the midplane,
where the magnetic field energy is $\sim 10^{-2}$ times the combined
gas and radiation pressure (see also \citealt{tur04}).
Likewise, in all cases the upper layers
of the disk are magnetically-dominated, and the photosphere lies
within this region.  Unlike the earlier simulations, in these two
the box-averaged radiation pressure is $\sim 10$ times
greater than the gas pressure at all times.

      Most importantly, in both cases, the energy content of the box
undergoes order unity fluctuations over timescales of many tens of
orbits, but these fluctuations have no long-term trend.  This result
directly challenges the prediction made in \citet{sha76} that radiation
pressure-dominated disk segments should be thermally unstable.
Note, however, that the prediction by \citet{lig74} of inflow
instability in radiation-dominated disks remains to be investigated.

In our simulations, the time-averaged dissipation rate in the disk body
is roughly equal to the characteristic value $c\Omega^{2}/\kappa_{\rm es}$,
the rate at which diffusive radiation flux maintains hydrostatic balance
\citep{sha76}.  However, in some places ($|z|\sim H$), the local dissipation
rate  exceeds the characteristic value by more than $30\%$.  If this
heat went into radiation flux, hydrostatic balance would be disrupted.
We find that the excess is exactly compensated by radiative advection
associated with an acoustic breathing mode.  At the same time, mechanical
work done on the fluid by the shearing boundaries in excess of the local
dissipation rate is transported outward by Poynting flux and radiation
pressure work associated with the breathing mode.  Both the dissipated energy 
carried by radiation advection and (non-dissipated) energy carried by
Poynting flux and radiation pressure work are finally deposited and dissipated
at $|z|\sim 2H$, and from there all the way through the rest of the structure
radiation diffusion overwhelms all other energy fluxes.

      To explain the thermal stability of radiation-dominated disks, we argue
that the comparability
of stress and pressure inferred from dimensional analysis (which
underlies the $\alpha$-model, and the prediction of instability)
is due to
dissipation of magnetic turbulence (which produces the stress) providing
the heat that is then transformed into radiation pressure.  Consequently,
fluctuations in magnetic energy drive fluctuations in pressure, and
not (as has been commonly assumed) the other way around.  Our claim
is supported by two lines of evidence: First, crosscorrelation analysis
of simulation data demonstrates that magnetic fluctuations {\it lead}
radiation energy fluctuations by 5--15 orbits, a little less than a thermal
time.  Pressure fluctuations cannot, then, drive magnetic fluctuations.
Second, we constructed a simple model realizing this picture, and this
model reproduces two other important features observed in the simulation data:
The system undergoes thermal fluctuations closely resembling in
amplitude and timescale those seen in the simulations.  In addition,
although {\it no} correlation between magnetic energy and radiation
pressure is built into the model, thermal balance automatically
creates one after the fact.  Thus, correlations between stress
and pressure are due to the dissipation of magnetic energy
supplying thermal energy, not to the pressure defining a characteristic
scale for the stress.

A logical consequence of this point of view, in which stress
determines pressure, rather than the other way around, is that
the fundamental independent variables are surface density
and orbital frequency.  That the orbital frequency is independent
of disk parameters is obvious, so long as the disk mass is small
compared to the central mass.  So long as the inflow timescale
is long compared to the thermal timescale, the surface density
must likewise be regarded as an independent parameter with
respect to thermal and dynamical fluctuations.   These two
independent parameters, through the intertwined
and nonlinear processes of MHD instability, tapping the
energy reservoir of orbital shear, magnetic dissipation,
thermal radiation, and radiative diffusion, with all of
these taking place under conditions of vertical (as
well as radial) gravitational confinement, combine to
determine the magnetic field strength, both its mean saturation level
and its fluctuations.  Orbital shear fixes the stress
exerted by this field, while the turbulent cascade sets the
dissipation rate.  These two are not entirely separate, of course,
as the time-averaged accretion rate is equivalent to either one.
The pressure follows from the heating
rate, as regulated by photon diffusion, and, in turn, closes
the loop by determining the disk thickness.  Despite all
these complications, at bottom, everything is still determined
by only two variables, surface density and orbital frequency.

We are grateful to Shane Davis, Jim Stone, and Neal Turner for very useful
discussions and comments.
This work was partially supported by NSF Grant AST-0507455 and
NASA ATP Grant NNG06GI68G (JHK) and NSF Grants AST-0307657 and
AST-0707624 (OMB).  The computations were performed on the SX8 at the Yukawa
Institute for Theoretical Physics of Kyoto University and the VPP5000 at
the Center for Computational Astrophysics of the National Astronomical
Observatory of Japan.

\newpage
\appendix
\section{Numerical methods for solving the matter-radiation energy exchange
equations}

To solve the equations for matter energy density $e$ (eqn.~\ref{eq:mattenergy})
and radiation energy density $E$ (eqn.~\ref{eq:radenergy}), we follow our
usual system of operator-splitting.  Because energy removed from one of
these reservoirs can go directly into the other, we treat the operator-split
segments in pairs.

\subsection{Free-free absorption}

First we group together the work done by matter on radiation and the
energy exchanged by free-free absorption and emission:
\begin{eqnarray}
 \frac{\partial e}{\partial t} &=& - (\nabla\cdot\bm{v})p - (4\pi B - cE)\bar{\kappa}_{\rm ff}^{\rm P}\rho\\
 \frac{\partial E}{\partial t} &=& - \nabla\bm{v}:\mathsf{P} +  (4\pi B - cE)\bar{\kappa}_{\rm ff}^{\rm P}\rho.
\end{eqnarray}
To solve these equations implicitly, we follow the method described in
\S~4.3 in \cite{tur01}, where the following quartic for
$x\equiv e^{n+1}$ is derived (the superscript is the time-step index):
\begin{equation}
x^{4}+c_{2}x + c_{1} = 0
\label{eq:quartic}
\end{equation}
where
\begin{eqnarray}
c_{1}&\equiv&-\frac{(1+a_{3}+a_{2})e^{n}+a_{2}E^{n}}{a_{1}(1+a_{3})}, \\
c_{2}&\equiv&\frac{(1+a_{4})(1+a_{3}+a_{2})}{a_{1}(1+a_{3})}, \\
a_{1}&\equiv&4\bar{\kappa}_{\rm ff}^{\rm P}\sigma_{\rm B}\left(\frac{\mu(\gamma-1)}{\mathcal{R}\rho^{n}}\right)^{4}\Delta t,\\
a_{2}&\equiv&c\bar{\kappa}_{\rm ff}^{\rm P}\Delta t,\\
a_{3}&\equiv&\frac{(\nabla\bm{v}:\mathsf{P})^{n+1}}{E^{n+1}}\Delta t,\\
a_{4}&\equiv&(\gamma-1)(\nabla\cdot\bm{v})\Delta t.
\end{eqnarray}
Here $\mathcal{R}$ and $\sigma_\mathrm{B}$ are the gas constant and the
Stefan-Boltzmann constant, respectively.  Once the quartic (\ref{eq:quartic})
is solved for $x=e^{n+1}$, $E^{n+1}$ is computed as
$(E^{n}+e^{n} - (1+a_{4})x)/(1+a_{3})$.

Hereafter we assume that $(1+a_{3}) = 
(E^{n+1} + (\nabla\bm{v}:\mathsf{P})^{n+1}\Delta t)/E^{n+1} > 0$ and
$(1+a_{4}) = (e^{n+1} + (\nabla\cdot\bm{v})p^{n+1}\Delta t)/e^{n+1} > 0$.  This
assumption is valid whenever $\Delta t$ is small enough for the changes in the
energy densities by photon damping and gas compression during the time-step
to be smaller than the energies themselves.  Then $c_{1} < 0$ and $c_{2} > 0$
because $a_{1} > 0$ and $a_{2} > 0$. The signs of $c_{1}$ and $c_{2}$ are crucial
in the following.\footnote{In our actual simulations, we exclude the gas
compression term and solve for its effect separately, which means $a_{4}=0$,
but $(1+a_{3}) > 0$ is always true.  If the photon damping term is also
treated separately ($a_{3} = 0$), it is always guaranteed that $c_{1} < 0$
and $c_{2} > 0$.}

\subsubsection{Solving the quartic (\ref{eq:quartic}) by Ferrari's formula}

The quartic ($\ref{eq:quartic}$) can be solved by Ferrari's formula. First we
rewrite the equation as
\begin{equation}
x^{4} = -c_{2}x - c_{1}.
\end{equation}
Adding $(tx^{2}+t^{2}/4)$ to both sides leads to a new quartic:
\begin{equation}
\left(x^{2}+\frac{t}{2}\right)^{2} = -c_{2}x - c_{1} + tx^{2} + \frac{t^{2}}{4}.
\label{eq:newquartic}
\end{equation}
If we could find a value $t$ such that
\begin{equation}
-c_{2}x - c_{1} + tx^{2} + \frac{t^{2}}{4} = \left(\sqrt{t}x-\frac{c_{2}}{2\sqrt{t}}\right)^{2},
\label{eq:cubic}
\end{equation}
we could rewrite equation~(\ref{eq:newquartic}) as
\begin{equation}
\left(x^{2}+\frac{t}{2}\right)^{2} = \left(\sqrt{t}x-\frac{c_{2}}{2\sqrt{t}}\right)^{2}.
\end{equation}
An equation in this form can be solved readily, giving the four roots
\begin{equation}
x_{1,2}=\frac{\pm\sqrt{t}+\sqrt{-t\mp(2c_{2})/\sqrt{t}}}{2},
\quad x_{3,4}=\frac{\pm\sqrt{t}-\sqrt{-t\mp(2c_{2})/\sqrt{t}}}{2}.
\end{equation}
Since $c_{1} < 0$ and $c_{2} > 0$ as discussed above, there is only one
positive solution for $x$:
\begin{equation}
x=\frac{-\sqrt{t}+\sqrt{-t+(2c_{2})/\sqrt{t}}}{2}.
\end{equation}
If we can assume that $t > 0$, this solution is real.  We next prove that this
is a good assumption.

\subsubsection{Solving the cubic (\ref{eq:cubic}) by Cardano's formula}

Equation~$\ref{eq:cubic}$ can be rewritten as a cubic for $t$:
\begin{equation}
t^{3}+3pt+q = 0,
\label{eq:newcubic}
\end{equation}
where $p\equiv-4c_{1}/3$ and $q\equiv-c^{2}$. Here $p>0$ and $q<0$
because $c_{1} < 0$ and $c_{2} > 0$.  The single positive solution of
this cubic can be obtained via Cardano's formula as follows:
\begin{equation}
t = \sqrt[3]{\alpha_{+}} + \sqrt[3]{\alpha_{-}},
\end{equation}
where
\begin{equation}
\alpha_{\pm} \equiv \frac{-q\pm\sqrt{q^{2}+4p^{3}}}{2}.
\end{equation}
Note that $t>0$, as required.

\subsection{Compton scattering}

Next consider the terms describing Compton scattering.  We rewrite
them in the form
\begin{eqnarray}
&&\frac{\partial E}{\partial t} =  
\left(\frac{4\sigma_{\rm es}}{\mu_{\rm e}m_{\rm e}m_{\rm p}c}\right)E
\left((\gamma-1)\mu m_{\rm p}e - k_{\rm B}\rho\left(\frac{E}{a}\right)^{1/4}
\right) \\
&&\frac{\partial E}{\partial t} + \frac{\partial e}{\partial t} = 0.
\end{eqnarray}
Here $\sigma_{\rm es}$, $m_\mathrm{p}$, and $m_\mathrm{e}$ are the Thomson
cross section, the mass of the proton, and the mass of electron, respectively.
We assume that the mean molecular weights $\mu$ and $\mu_{\rm e}$ are $0.61$
and $1.2$ respectively.  We solve these equations implicitly:
\begin{eqnarray}
&&\frac{E^{n+1}-E^{n}}{\Delta t} = \left(\frac{4\sigma_{\rm es}}
{\mu_{\rm e}m_{\rm e}m_{\rm p}c}\right)E^{n+1}
\left((\gamma-1)\mu m_{\rm p}e^{n+1} - k_{\rm B}\rho^{n}
\left(\frac{E^{n+1}}{a}\right)^{1/4}\right) \\
&&\frac{E^{n+1}-E^{n}}{\Delta t} + \frac{e^{n+1}-e^{n}}{\Delta t} = 0.
\end{eqnarray}
Eliminating $e^{n+1}$ from this pair of equations, we find a nonlinear
equation for $x\equiv(E^{n+1}/E^{n})^{1/4}$:
\begin{equation}
x^{4}\left(x^{4} + Ax + \left(B - 1 -\frac{e^{n}}{E^{n}}\right)\right) - B = 0,
\label{eq:nonlinear}
\end{equation}
where $A\equiv (k_{\rm B}\rho^{n}(E^{n}/a)^{1/4})/
((\gamma-1)\mu m_{\rm p}E^{n})$ and $B\equiv (\mu_{\rm e}m_{\rm e}c)/
(4\mu(\gamma-1)\sigma_{\rm es}E^{n}\Delta t)$.
We solve the nonlinear equation~\ref{eq:nonlinear} by the Newton-Raphson method.
That gives us the energies for the next step, $E^{n+1} = x^{4}E^{n}$ and
$e^{n+1} = (E^{n}+e^{n}) - E^{n+1}$. 

\end{document}